# Remote Key Establishment by Mode Mixing in Multimode Fibres and Optical Reciprocity

**Yaron Bromberg, Brandon Redding, Sebastien M. Popoff[†] and Hui Cao***
*Department of Applied Physics, Yale University, New Haven, Connecticut 06520 USA*

*hui.cao@yale.edu

[†] Current address: CNRS - LTCI Telecom ParisTech, 46 rue Barrault, 75013 Paris, France

**Disorder and scattering in photonic systems have long been considered a nuisance that should be circumvented. Recently, disorder has been harnessed for a rapidly growing number of applications, including imaging, sensing and spectroscopy. The chaotic dynamics and extreme sensitivity to external perturbations make random media particularly well-suited for optical cryptography. However, using random media for distribution of secret keys between remote users still remains challenging, since it requires the users have access to the same scattering system. Here we utilize random mode mixing in multimode fibres to generate and distribute keys simultaneously. Fast fluctuations in the fibre mode mixing provide the source of randomness for the key generation, and optical reciprocity guarantees that the keys at the two ends of the fibre are identical. We experimentally demonstrate the scheme using classical light and off-the-shelf components, opening the door for cost-effective key establishment at the physical-layer of fibre-optic networks.**

# 1. Introduction

Complex photonic systems, such as scattering media, chaotic cavities, aperiodic photonic crystals and biological tissue, are comprised of a large number of spatial, spectral, temporal and polarization degrees of freedom. The strong coupling of these degrees of freedom provides exceptional opportunities for numerous applications, e.g., the spatial-temporal coupling was utilized for dynamic light scattering and diffusive wave spectroscopy[1,2], the spatial-spectral coupling for spectroscopy[3] and imaging[4,5], and the spatial-polarization coupling for polarimetry[6]. For optical cryptography, complex random media has been utilized for several cryptographic tasks, including authentication[7,8], identification[9], encryption[10], random number generation[11,12], and secure key storage[13]. An optical fibre that supports hundreds or thousands of guided modes can also be considered as a complex photonic system. Random mode mixing arises naturally due to local index inhomogeneities and cross section variations, producing speckle patterns at the output of the fibre. Since the mixing depends on ambient temperature fluctuations and mechanical strains[14], the output speckle pattern is extremely sensitive to environmental perturbations and therefore constantly changes in time. This poses a serious challenge for many applications such as telecommunication and imaging, as the transmitted information quickly gets scrambled. Here, we take advantage of the random chaotic fluctuations in a long multimode fibre to generate random keys. Most importantly, the remote users at the two ends of the fibre can share identical copies of the keys, by virtue of the optical reciprocity principle. The keys, which are constantly updated due to intrinsic fluctuations of the fibre, can then be used to encode and decode the information being sent over a standard unsecure communication channel.

Our method of distributing keys at the physical-layer of communication networks can only be hacked in real-time, in contrast to the computational keys that are widely used in today's telecommunications. Compared to the quantum key distribution (QKD), our scheme uses classical light and off-the-shelf fibre components, and is therefore much simpler and can be easily integrated with current communication networks. While other classical key distribution approaches, e.g., the ones using chaotic lasers[15], require a precise tuning of the system parameters, ours is alignment-free and naturally robust, making it especially attractive for real-world applications. Recently, several key distribution methods based on single mode fibres have been developed[16,17], however, they are prone to eavesdropping because of the simplicity of the single mode fibre. In our approach, the complexity of a multimode fibre forces an adversary to simultaneously measure all fibre modes to extract information, and the measurements must be done quickly to track the rapid fluctuations of the fibre. The additional noise introduced by such exhaustive measurements sets the security of the key.

## 2. Proof of principle demonstration

Let us consider two users, Alice and Bob, who simultaneously couple laser light of identical frequency into both ends of a multimode fibre (Fig. 1). As an example, we assume the input beam from Alice (Bob) has a well-defined wavevector $\vec{k}_A$ ($\vec{k}_B$). In the presence of strong mode mixing, by the time the light exits the fibre all the guided modes are excited. Thus, at the far-field of the output facet a speckle pattern emerges, with no trace of the input wavevector [Fig. 1(a)]. The patterns Alice and Bob observe result from exactly the same mode mixing and phase shifts across the fibre, but in a reversed order. Since the order of the coupling events is not interchangeable, the output patterns at the opposite ends are different even when the input channels of Alice and Bob are the same ($\vec{k}_A = \vec{k}_B$). Nevertheless, there is a unique pair of output channels that will be perfectly correlated. Optical reciprocity guarantees that the intensity measured by Alice at the output channel $-\vec{k}_A$, is identical to the intensity measured by Bob in the output channel $-\vec{k}_B$ [Fig. 1(b)]. Remarkably, not only does this hold even when the two input channels are not identical ($\vec{k}_A \neq \vec{k}_B$), but Alice and Bob do not need to know which channel the other user couples light into at the opposite end of the fibre. All they need to do is measure the output intensity from the same channel that they have coupled light into. As the environmental conditions of the fibre change, the speckle patterns and the intensities measured in the $-\vec{k}_A$ and $-\vec{k}_B$ channels will fluctuate. However, since reciprocity holds for any configuration of the fibre, the intensities measured in the $-\vec{k}_A$ and $-\vec{k}_B$ channels will remain perfectly correlated [Fig. 1(c)], as long as the fibre remains static during the time it takes the light to traverse the it. This is demonstrated experimentally in Fig. 1(d). Since the input and output channels are specified not only by the wavevector, but also by the polarization, we have placed linear polarizers at both ends of the fibre. Note, however, that the two polarizers do not need to select the same polarization state, but can be oriented at any arbitrary angle. To emulate changes in the environmental conditions of long fibres, the fibre was constantly shaken while speckle patterns were recorded. The synchronized fluctuations of the intensities measured by Alice and Bob allow them to share a common random signal from which they can extract a key.

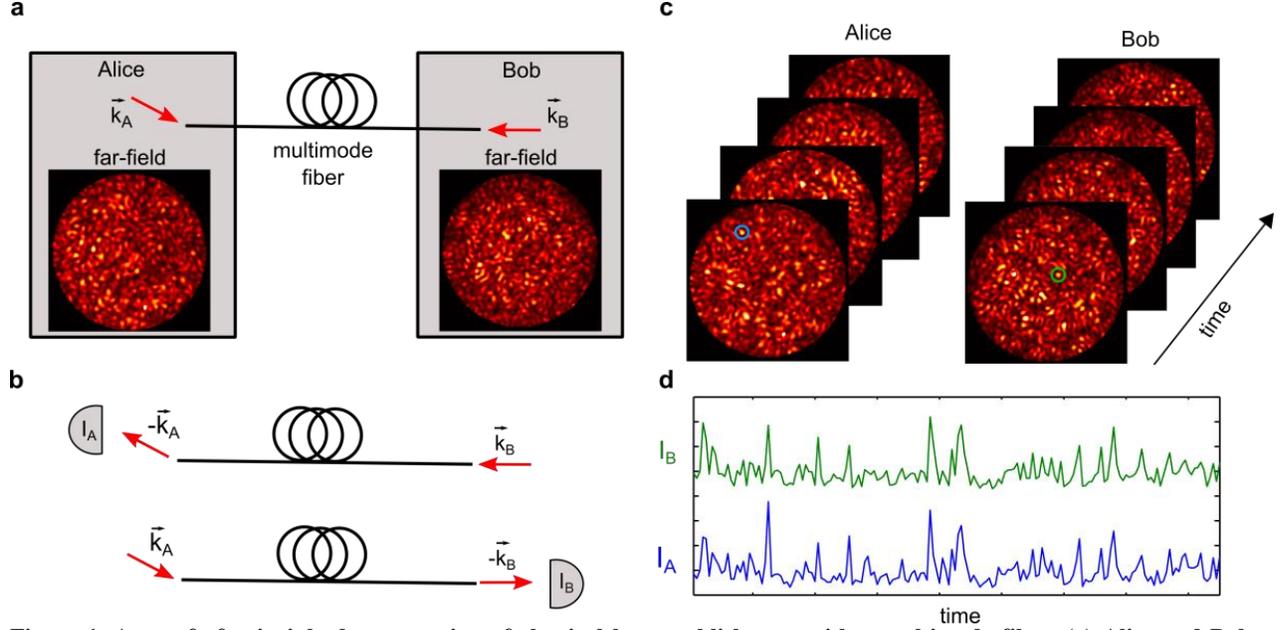

**Figure 1. A proof of principle demonstration of classical key establishment with a multimode fibre. (a)** Alice and Bob couple a laser beam to the multimode fibre with wavevectors $\vec{k}_A$ and $\vec{k}_B$ respectively. Due to mode mixing in the fibre, speckle patterns appear at the output. The fibre is not perfectly symmetric with respect to its centre, therefore Alice and Bob measure two different patterns. **(b)** According to the optical reciprocity theorem, when the positions of a detector and a source are interchanged, the intensity measured by the detector in the two different configurations ($I_A$ and $I_B$) is identical. Thus, at the channels that correspond to $-\vec{k}_A$ and $-\vec{k}_B$, Alice and Bob measure exactly the same intensity. **(c)** Due to changes in the environmental conditions of the fibre, the output speckles constantly fluctuate. Nevertheless, at channel $-\vec{k}_A$ (blue circle) and $-\vec{k}_B$ (green circle) the intensity is always correlated, as verified experimentally in **(d)** with a 6-meter long step-index fibre (core diameter = 105 μm, numerical aperture = 0.22) that supports $M \approx 4000$ guided modes. The probe laser wavelength is $\lambda = 640$ nm.

While the above demonstration is performed in the wavevector (k) space, the reciprocity principle is not restricted to the wavevector space, and Alice and Bob can use channels in other spaces, e.g., the guided-mode space or the position space. Moreover, either of them can use channels in a different space, without knowing which space is used by the other. The position space is convenient for fibre-network applications, because Alice and Bob can simply couple light to each end of the multimode fibre via single mode fibres. The single mode fibres, together with an in-line fibre polarizer, automatically guarantee that the illumination and detection are conducted in the same channels, enabling an all-fibre, alignment-free configuration which is compatible with optical fibre networks. Furthermore, in contrast to QKD, our method does not have to operate at the single photon level and can easily be implemented using standard telecommunication lasers. Figure 2(a) is a schematic of an all-fibre set-up we built with off-the-shelf elements. Using this setup, we demonstrated that, the intensities measured by Alice and Bob exhibit a correlation of 0.99 [Fig. 2(b)]. Such a high degree of correlation allows digitization of the analogue signal, by associating a bit value '1' to all the intensities above the median, and '0' to all the intensities below the median. After digitization, the correlation between the stream of bits Alice and Bob obtain is 0.90, and the bit error rate is 0.05 [Fig. 2(c)], which can be further reduced with error correction protocols[18]. The key rate is determined by the rate of the fibre fluctuations, which in our case was only 20 Hz, since we perturbed just 5 meters out of a 1 km long fibre. However, the reported fluctuation rates of long-haul fibres in optical networks are much higher, typically in the range of 1-100 KHz[19,20], enabling high key generation rates. The key rates can be further increased via parallelization, for example using wavelength division multiplexing (WDM).

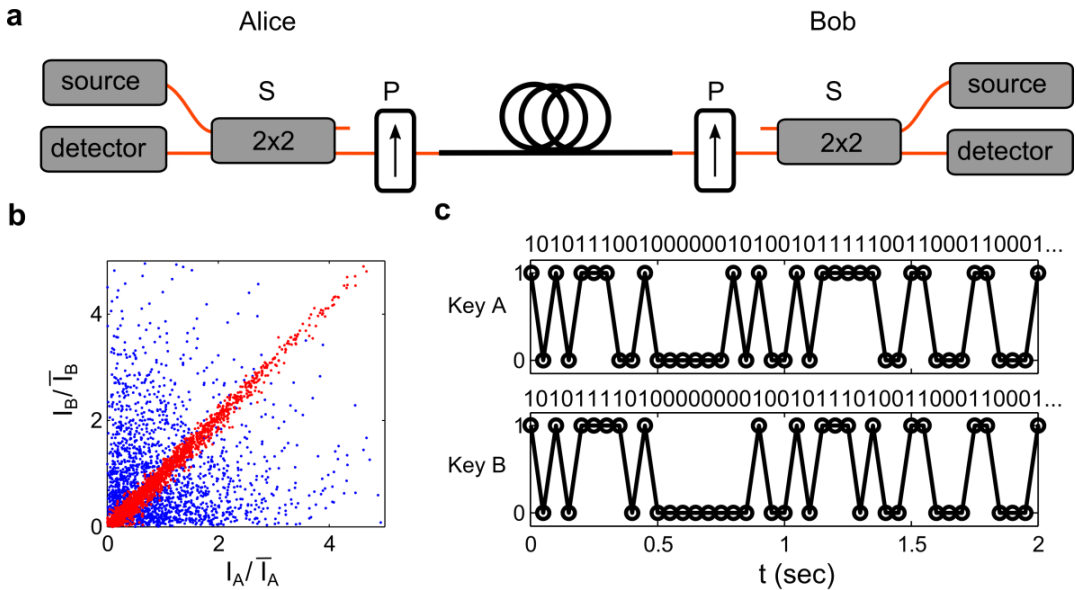

**Figure 2. Telecommunication compatible implementation of remote key distribution. a) Each node contains a laser source, a photodetector, a 2x2 splitter (S) and an in-line fibre polarizer (P). In our experiment a single laser source operating at the telecommunication C-band ($\lambda$=1550nm) is used with a splitter (not shown) to mimic two sources at both ends of the fibre. The laser light is coupled to the multimode fibre (black) via a single mode fibre (orange) to ensure that the illumination and detection channels are identical at either end of the fibre. The multimode fibre is a 1 km long graded-index fibre with a core diameter of 62.5 μm, supporting about $M \approx 100$ guided modes. The environmental changes on the fibre are induced by shaking a 5-meter portion of the fibre (out of the 1 km spool). (b) Scatter plot of the intensity measured by Alice versus the intensity measured by Bob, exhibiting a correlation of 0.99 (blue dots). To verify that the high level of correlation does not result from fluctuations in the total transmitted light, we added a free space beam slitter between the single mode fibre and the multimode fibre at Bob's end (not shown), and measured the intensity at an arbitrary position across the beam, yielding a correlation of 0.01 (red dots). (c) A fraction of the digitized key, showing a raw bit rate of 20 Hz, and a bit error rate of 0.05.**

## 3. Eavesdropping and security analysis

Next we analyse possible attacks by an eavesdropper, Eve. The attacks fall under two categories, passive ones where Eve only probes the light that is transmitted between the users, and active ones where Eve can also inject light into the system and use additional modulators to deceive Alice and Bob. In this section we focus on passive attacks and briefly discuss active attacks at the end. We specifically consider the so-called beamsplitter attack, in which Eve places a beamsplitter at some intermediate point along the fibre.

Eve can then measure the speckle intensity patterns incident at the beamsplitter, travelling from Alice to Bob and from Bob to Alice [Fig. 3(a)]. These two patterns are uncorrelated since they have travelled through two different sections of the multimode fibre and experienced different mode mixing. Hence, none of the spatial channels measured by Eve will be correlated with the signals Alice and Bob measure. We experimentally emulated a beamsplitter attack and confirmed that the intensities in all the channels measured by Eve are uncorrelated with the intensity measured by Alice and Bob [Fig. 3(b)]. Note that, as in any cryptographic system, we assume the nodes of Alice and Bob are in a secure area which Eve cannot access. Alice and Bob can therefore make sure that the sufficient mode mixing is introduced to the portion of the multimode fibre within the secure area so that when the light exits the secure area, all channels are excited and the intensity measured by Eve in any channel will be uncorrelated with the signal sent by Alice and Bob.

The random mode mixing along the fibre prevents Eve from extracting the key via direct intensity measurements. However, a beamsplitter attack provides Eve access not only to the amplitude, but also to the phase of the counter-propagating fields. Even in the presence of strong mode mixing, the complex fields at the output port of Eve's beamsplitter carry information about the key. To illustrate this, let us denote the complex field distribution at the output port of Eve's beamsplitter from Alice by $E_{A,p}(x,y)$, where $p = $ H,V denotes the polarization state. Since at a single frequency phase conjugation corresponds to the time-reversal operation, a complex conjugate field $E^*_{A,p}(x,y)$ will send the light back to Alice's channel. Thus the projection of $E^*_{A,p}(x,y)$ onto field distribution that Eve measures at the output port from Bob $E_{B,p}(x,y)$, gives the fraction of the light that actually reaches Alice's channel. Hence, Eve can extract the key by computing the complex overlap of the field distributions:

$$I = \left|\sum_{p=H,V} \iint dxdy E_{A,p}(x,y) E_{B,p}(x,y)\right|^2. \qquad (1)$$

While this is theoretically possible, such a full-field beamsplitter attack requires an extremely complex measurement apparatus. For *every* spatial channel, Eve needs a local oscillator to coherently detect the two quadratures of the field, at two orthogonal polarizations, for the two counter propagating fields. These can add up to thousands of channels that need to be measured faster than the fluctuation rate of the field. Despite of recent advances in the spatial division multiplexing (SDM) in multimode fibres[21], such a measurement is still out of reach of current technology. Furthermore, even if the technology for rapid full-field characterization will become available in the future, the excess noises induced by Eve's measurements will impose a fundamental limit on the amount of information that she can extract.

To analyse the fundamental noise associated with the measurements of Alice, Bob and Eve, we assume the laser light is in a coherent state with a large mean photon number. In this limit, Alice and Bob can perform a direct intensity measurement which is limited only by shot-noise[22]. Eve, however, needs to first coherently detect the field distributions $E_{A,p}(x,y)$ and $E_{B,p}(x,y)$, then compute the intensity $I$ according to Eq. (1). Noise is added to her signal at two levels (See Supplementary Section 2). First, for coherent states, the standard deviation in the simultaneous measurement of the two field quadratures is $\sqrt{2}$ larger than that of a single quadrature measurement[23]. Thus, the standard deviation of the intensity computed from the sum of the squares of the quadratures is $\sqrt{2}$ larger than that of a direct, shot-noise limited, intensity measurement. A second source of noise in Eve's reconstruction is the multiplication of fields $E_{A,p}(x,y)$ and $E_{B,p}(x,y)$ in the right hand side of Eq. (1). Due to error propagation, multiplication of two random variables increases the standard deviation of the product by another factor of $\sqrt{2}$. Thus, if Eve detects $N_E$ photons per channel, her signal-to-noise ratio (SNR) is $\sqrt{N_E/4}$, whereas the SNR of Alice and Bob for a shot-noise limited detection of $N$ photons per channel is $\sqrt{N}$. We therefore conclude that as long as the number of photons per channel that Eve measures is a factor of 4 (6 dB) less than the number of photons that Alice or Bob measure, Eve's SNR will be lower than that of Alice and Bob.

Previous information theoretic security analysis[24] has proven that the security of key distribution at the physical-layer is guaranteed as long as the legitimate users (Alice, Bob) have access to a common source of randomness, through channels that are less noisy than the channel of the eavesdropper (Eve). Thus, the superior SNR of their measurements enables Alice and Bob to generate a secure key, provided that they also have access to an unsecure public channel for privacy amplification protocols. We conduct a quantitative analysis based on mutual information, and numerically calculate the secure key fraction (K), defined by the ratio of the secure bit rate and the raw bit rate (See Supplementary Section 2 for further details). Figure 3(c) plots the secure key fraction $K$ as a function of the reflectivity of Eve's beamsplitter, confirming that as long as Eve measures less than 80% of the photons in the fibre, security is guaranteed. Obviously, if Eve takes 80% of the total transmitted power, Alice and Bob would detect her presence. Moreover, Eve *cannot* compensate for the light she removes from the fibre, because for shot-noise limited signals, amplification cannot improve the SNR[25] and Alice and Bob will therefore notice an increase in the noise they measure.

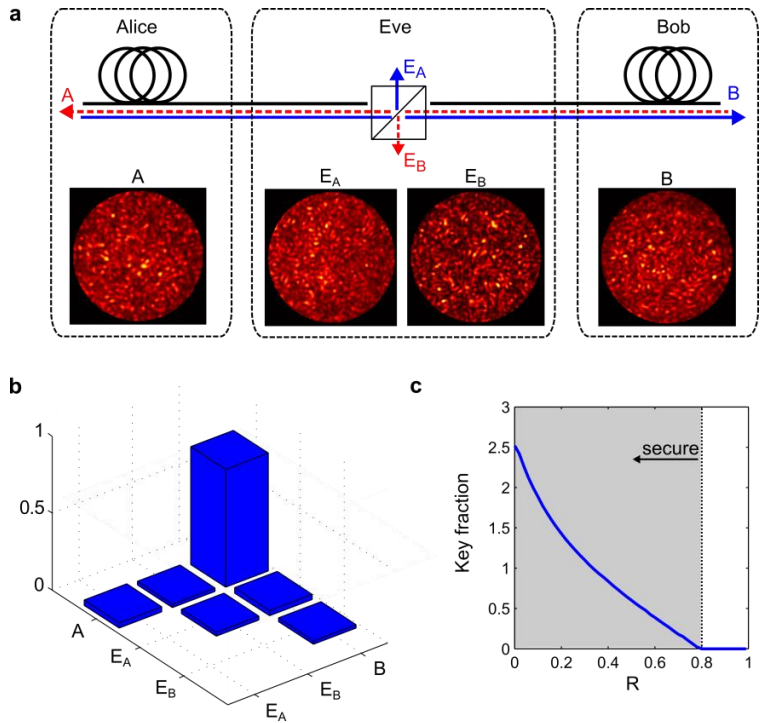

Figure 3. Experimental emulation of a direct beamsplitter attack on the system shown in figure 1. (a) Eve picks up a fraction of the light propagating between Alice and Bob by intersecting the fibre with beamsplitter. The light which arrives to $E_A$ and $E_B$ passes through different parts of the fibre and therefore experiences completely different mode mixing. (b) The cross-correlations of measured intensities between the speckle grains from the four patterns A, B, $E_A$ and $E_B$. The cross-correlation between A and B corresponds to the input modes. Due to strong mode mixing, all the speckles in patterns $E_A$ and $E_B$ are equivalent, and two of them are selected from $E_A$ and $E_B$ for the other five cross-correlations. (c) Secure key fraction K as a function of the beamsplitter reflectivity R, for a simulated full-field attack by Eve. For R<0.8, a secure key is established.

Finally, we note that thus far we have not considered the fibre transmission loss. Transmission loss benefits Eve, since the light she detects experiences less loss than the light which Alice and Bob detect. The mutual information analysis illustrates that the optimal position for Eve to place the beam splitter is at the midpoint of the fibre. Then, Alice and Bob can establish a secure key only if the transmission from the midpoint of the fibre to its ends is higher than -6dB. For typical fibres with transmission loss of 0.2db/km[26], this corresponds to a maximum fibre length of 60 km.

Below we briefly discuss a few potential active attacks. An active Eve can try to perform the so-called man-in-the-middle attack, where she blocks the light that Alice and Bob send, and couples her own light into the fibre. However, by comparing part of the keys they obtain over a public channel, Alice and Bob can immediately notice that their keys are no longer correlated, and stop the communication. If Eve tries to inject light to the fibre without blocking the light that Alice and Bob send, they can notice a reduction in the contrast of the intensity fluctuations they measure. To prevent Eve from inducing correlations by modulating the total intensity of the light passing through the fibre, Alice and Bob can measure the total intensity that arrives to their end of the fibre, and verify that it is not modulated.

**4. Discussion and conclusions**

Our approach to spread the transmitted signal over multiple spatial channels bares similarity with key distribution methods in the wireless domain, which rely on the scattering and fading of radio-frequency waves[27]. In the optical domain, a recent work demonstrated key distribution by free space propagation through turbulent media, but in this configuration the legitimate users measure the reciprocal phase using a complex detection apparatus[28]. Our scheme is much simpler and more robust as it relies on a direct intensity measurement. Compared to the previous key-distribution schemes based on single-mode fibres[16,17], the complexity of the multimode fibre forces the hacker to use a significantly more complex detection apparatus and perform additional computational processing. The asymmetry between the complex detection of the adversary and the simple direct detection of the legitimate users imposes an excess noise in the adversary's measurements, which the legitimate users can utilize via privacy amplification protocols to distil a secure key[29].

Finally, we further discuss the maximal fibre length. In addition to the 60 km limit due to the transmission loss mentioned above, two additional factors affect the fibre length. First, since the fibre has to be static during the propagation time, the maximal length depends on the raw key rate. For a raw key rate of 1 KHz for example, the maximal fibre length is about 200 km. The second factor is the modal bandwidth $\delta$, also known as the spectral correlation width, which decreases monotonically with the fibre length. The modal bandwidth sets the maximal possible frequency mismatch between the lasers of Alice and Bob. The typical value of $\delta$ is about 500 MHz·km, which for 200 km long fibres gives 2.5 MHz. We note that the bandwidth of the lasers, $\Delta$, can exceed $\delta$; but the contrast of the fluctuations will be reduced by a factor of $\sqrt{\Delta/\delta}$. If fluctuations on the order of 5% could be used to generate the key, the available optical bandwidth for a 200 km fibre would be approximately 1 GHz.

In conclusion, we have developed a key establishment protocol that relies on classical light and off-the-shelf telecommunication components. We experimentally demonstrate the method using an all-fibre and alignment-free system, which is compatible with optical fibre networks. The security of our method relies on the complexity of multimode fibres, and on the fundamental noise limits of coherent photodetection. While this work has focused on classical light, we notice an analogy between our method and continuous variable quantum key distribution (CVQKD), as described in Supplementary Section 2. This analogy may allow a future extension of our security analysis from the specific attack considered here to the more general security proofs that were recently developed for CVQKD[30]. The analogy to CVQKD may also help developing a new protocol for overcoming the 60 km distance limit discussed above, similar to the way CVQKD has been extended beyond the 3dB loss limit using postselection[31]. In addition, spreading quantum light over multiple spatial channels may open new opportunities also for QKD, in the spirit of the recent proposals to utilize multiple temporal channels for advanced QKD protocols[32].

# Remote Key Establishment by Mode Mixing in Multimode Fibres and Optical Reciprocity: Supplementary Information

## 1. Comparison with other key distribution methods

Our scheme is inspired by the on-going effort to distribute keys between remote users at the physical-layer of communication networks. To date, the security of telecommunication networks depends on computational security, which is based on unproven assumptions that some computational tasks cannot be computed efficiently using a classical computer[S1]. However, an adversary can tap the communication channel, save all the encrypted bits, and decrypt the information once a quantum computer, or a new computational paradigm, becomes available. In contrast, key distribution at the physical-layer can be compromised only if it is hacked in real-time[S2]. Quantum key distribution (QKD) is one such physical-layer approach, which relies on the fundamental laws of quantum mechanics to guarantee that an eavesdropper cannot extract the key without being exposed. However, a quantum channel requires transmission and detection of quantum states of light, which is technically challenging and expensive, and thus has not been widely implemented yet.

Over the years several classical alternatives to QKD have been proposed. A common approach is to rely on a pre-established secret information, for example, the parameters of a chaotic laser[S3], or the configuration of the encoder and decoder[S4]. This approach typically requires accurate tuning by the end users so that their system parameters will match. More importantly, to guarantee security these settings have to be constantly updated, and therefore require an additional key distribution method for securely sharing the updated parameters. In an alternative approach, the entire channel between the users is turned to a giant fibre laser, with switchable cavity mirrors[S5]. From the lasing characteristics each user knows what mirror was used at the other end of the fibre, and utilize this information to generate a key. However, an eavesdropper can directly measure the reflectivity from the mirrors and extract the key, since the key is deterministically set by the configuration of the cavity mirrors. It is therefore essential that the key will not be determined by the settings of just a few elements, which the adversary can measure. Towards this end, it was recently demonstrated that the relative phase between two fluctuating single mode fibres can be used for generating a key[S6]. Since the phase difference between the two fibres is accumulated along the entire length of the fibres, it is significantly more challenging for an eavesdropper to measure. Nevertheless, by splitting each fibre and measuring the phase accumulated at each of the four segments, the eavesdropper can compute the total phase difference between the fibres and extract the key[S6]. In our multimode fibre configuration, an equivalent attack would require coherent detection of hundreds of spatial channels, simultaneously, featuring the complexity of multimode fibres. It forces the adversary to use a significantly more complex detection apparatus than the apertures of the legitimate users, and additional computational processing, which as we further discuss in the next section, sets the security of the key.

**2. Security analysis of full-field attack**

The fact that some information can leak to the eavesdropper is common to all physical-layer key distribution methods, including quantum key distribution[S7]. Nevertheless, security of the key can still be guaranteed provided that the amount of information that has leaked is smaller than the amount of information that is exchanged between the legitimate users, by processing the raw signals using privacy amplification protocols[S8]. The analysis of physical-layer systems is therefore based on notions from information theory that quantify the amount of information between the users. A central result in information theoretic security was derived by Csiszar and Korner[S9], who proved that the security of key distribution at the physical-layer is guaranteed as long as the legitimate users have access to a common source of randomness, through channels that are less noisy than the channel of the eavesdropper. Thus, we next perform a noise analysis of the full-field attack.

*Noise analysis of the full-field attack*

In this section we compare the signal-to-noise ratio (SNR) of the intensity reconstructed by Eve, with the SNR of the intensity measured by Alice and Bob. Let us denote by $A_m$ ($B_m$) the projection of the complex field $E_{A,p}(x,y)$ ($E_{B,p}(x,y)$), that gets from Alice (Bob) to Eve, onto the output channel $m = 1..2M$. Here $m$ denotes the output spatial and polarization channels at Eve's beamsplitter, and $M$ is the number of spatial channels the fibre supports. From Eq. (1) in the main text, Eve can reconstruct the intensity that is measured by Alice and Bob, by computing the intensity $I = |\sum_m A_m B_m|^2$. The signal reconstructed by Eve, however, will have an additional noise that originates from the noise in her measurements. We model the noise in the quadratures of each of the channels she measures by additive, uncorrelated, Gaussian random variables, with a variance $\sigma^2$. By plugging the random variables into the sum Eve computes and averaging over the noise realizations, we get the signal to noise ratio (SNR) of Eve's reconstructed intensity:

$$SNR_E = \sqrt{\frac{N_A N_B}{8M\sigma^2(N_A+N_B)}}, \quad (S1)$$

where $N_A$ ($N_B$) is the number of photons that get to Eve from Alice (Bob), and we assumed the high SNR limit $N_A, N_B \gg \sigma^2$. Assuming that Alice and Bob send the same number of photons per bit, $N_0$, then the average number of photos that arrive to Eve is $N_A = N_0 R e^{-\gamma(L/2+z)}$ and $N_B = N_0 R e^{-\gamma(L/2-z)}$, $\gamma$ is the fiber attenuation coefficient, $L$ is the fiber length, $z$ is the distance of Eve's beamsplitter from the fiber midpoint, and $R$ is the reflectivity of the beamsplitter. We thus get:

$$SNR_E = \sqrt{\frac{N_0 R e^{-\gamma L/2}}{8M\sigma^2} \frac{1}{2\cosh(\gamma z)}}. \tag{S2}$$

To maximize her SNR, Eve needs to place her beamsplitter at the midpoint of the fibre (z=0). The SNR is then given by:

$$SNR_E = \sqrt{\frac{N_E}{8\sigma^2}}, \tag{S3}$$

where $N_E = N_0 R e^{-\gamma L/2}/2M$ is the average number of photons per channel that get to Eve. For computing the fundamental limit on Eve's SNR, we need to evaluate the variance of the noise in the measurement of the two quadratures of the field. According to the quantum theory of photodetection, the fundamental noise associated with coherent detection of the two quadratures of a coherent state simultaneously is Gaussian noise with $\sigma^2 = 1/2$, and therefore $SNR_E < \sqrt{N_E/4}$. In contrast to Eve, Alice and Bob can perform a direct shot noise limited intensity measurement, and therefore their SNR is giving by $SNR_{A/B} = \sqrt{N}$, where $N = N_0(1-R)e^{-\gamma L/2}/2M$. We therefore conclude that as long as the number of photons per channel that Eve measures is a factor of 4 less than the number of photons that Alice or Bob measure, Eve's SNR will be lower than that of Alice and Bob.

*Mutual information analysis*

The noise in the signal reconstructed by Eve intuitively limits the amount of information she can extract. To obtain a quantitative relation between the amount of information obtained by Eve and the security of the key, we use the result of Csiszar and Korner, who proved that the key fraction, defined by the ratio of the secret key and the raw bit rate, is given by[S9]:

$$K = I(A,B) - \min[I(A,E), I(B,E)]. \tag{S3}$$

Here $I(A,B)$ is the mutual information between Alice and Bob, and $I(A,E)$ ($I(B,E)$) is the mutual information between Eve and Alice (Bob).

To study the effect of the excess noise on the key fraction, we performed a numerical simulation of Eve's attack and computed the mutual information of the simulated signals measured by Alice, Bob and Eve. The channel projections of the complex fields at Eve's output channel, $A_m$ and $B_m$, are simulated by circular complex Gaussian random numbers, and the intensity at the output channel of Alice and Bob is computed using $I_0 = |\sum_m A_m B_m|^2$. We next simulated Eve's noisy measurements by adding a Gaussian noise to the real and imaginary part of the complex fields $A_m$ and $B_m$ with variance $\sigma^2 = 1/2$, and computed Eve's noisy intensity. The noisy measurement of Alice and Bob are simulated by a Poissonian random number with mean $I_0$. We repeat this procedure for $10^6$ realizations of the fibre configuration (i.e. $10^6$ realizations of $A_m$ and $B_m$). Finally, we compute the mutual information between the simulated intensities of Alice, Bob and Eve, and using Eq. (S3) we obtain the secret key fraction K.

Figure S1 depicts the key fraction (K) as a function of the reflectivity of Eve's beam splitter, for different average numbers of photons per channel. As long as Eve measures less than 80% of the photons in the fibre, security is guaranteed. The 80% limit corresponds to the above observation that Eve needs to detect at least 6db more photons per channel than Alice and Bob, in order to hack the key. Obviously, if Eve tries to take 80% of the power, Alice and Bob will detect her presence. Moreover, she *cannot* compensate for the loss that Alice and Bob will experience, because for shot-noise limited signals, amplification cannot increase the SNR[S10]. Alice and Bob will therefore notice an increase in the noise they measure, and a reduction in their mutual information.

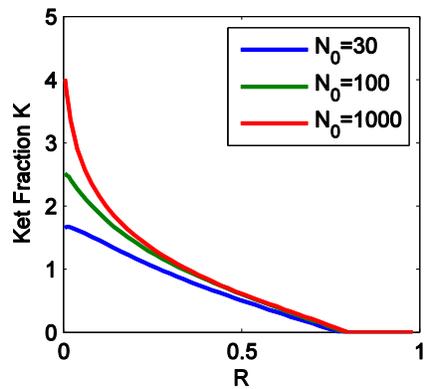

Figure S1. Numerical simulation of the secure key fraction K as a function of the reflectivity of Eve's beamsplitter, for different values of average number of photons per channel. Security is guaranteed for K>0, which is obtained for R<0.8.

*Analogy to continuous variable quantum key distribution*

Finally, we note that in this work we focused on analysing passive beam splitter attack. A promising route to extend the security analysis to more general attacks is via the analogy between our method and continuous variable quantum key distribution (CVQKD). In CVQKD Alice encodes a random key in the amplitude and phase of a coherent state using modulators, and sends it to Bob, who measures the two quadrature of the field using coherent detection[S11,S12]. Conceptually, one can think of our multimode fibre as the amplitude and phase modulator, which one of the users, say Alice, uses to send a random coherent state to Bob. Since Alice does not know the modulation which the multimode fibre has applied, she receives that information from the light Bob sends her. We encoded the key by the intensity measured by Alice and Bob, but similarly to CVQKD, the key can also be encoded by the quadratures of the fields. This analogy may allow a future extension of our security analysis, from the specific attack we considered here, to the more general security proofs that were recently developed for CVQKD[S13].